\documentclass[10pt,tightenlines,eqsecnum,floats,aps,amsmath,
amssymb,nofootinbib,prd,showpacs]{revtex4-1}
\usepackage{amsmath,amsthm,latexsym,amssymb,amsfonts}
\usepackage{subfig}
\usepackage{enumerate}
\usepackage{graphicx}
\usepackage{calc}
\usepackage{color}
\usepackage{vmargin}

\newcommand{\Hil}{{\mathcal H}}

\newcommand{\gora}[7] 
{
  \put(0,0){\qbezier(#1,#2)(#1,#4)(#3,#4)}
\put(0,0){\qbezier(#3,#4)(#5,#4)(#5,#6)}
\put(#3,#4){#7}
}
\newcommand{\pudla} 
{
\newsavebox{\pudlo}
\savebox{\pudlo}(2,1)[bl]{
\put(0,0){\line(1,0){1.5}}
\put(0,0){\line(0,1){0.8}}
\put(1.5,0){\line(0,1){0.8}}
\put(0,0.8){\line(1,0){1.5}}
\put(0.5,0.2){$P$}}
}

\begin{document}

\title{Operator Spin Foam Models}
\author{Benjamin Bahr${}^{1,2}$, Frank Hellmann ${}^3$,\\ Wojciech
Kami\'nski${}^3$,
Marcin Kisielowski ${}^3$,
Jerzy Lewandowski ${}^3$}

\affiliation{
${}^1$ DAMTP,
Centre for Mathematical Sciences, 
Wilberforce Road,
Cambridge CB3 0WA,
UK\\
${}^2$Albert Einstein Institute,
Muehlenberg 1,
14476 Golm,
Germany\\
${}^3$ Instytut Fizyki Teoretycznej, Uniwersytet Warszawski,
ul. Ho{\.z}a 69, 00-681 Warszawa (Warsaw), Polska (Poland)\\
}

\begin{abstract} \noindent{\bf Abstract\ }
The goal of this paper is to introduce a systematic approach to spin foams.
We define  operator spin foams, that is foams labelled by group representations
and operators, as the main tool. An equivalence relation we impose in the set of the operator spin foams allows to split the faces and the edges of the foams. The consistency with that relation requires introduction of the (familiar for the BF theory) face amplitude. The operator spin foam models are defined quite generally. Imposing a maximal symmetry leads to a family we call natural operator spin foam models.   This symmetry, combined with demanding consistency with splitting the edges, determines  a complete characterization of a general natural model.
It can be obtained by applying arbitrary (quantum) constraints on an arbitrary BF spin foam model. In particular, imposing suitable constraints on Spin(4) BF spin foam model is exactly the way we tend to view 4d quantum gravity, starting with the BC model and continuing with the EPRL or FK models. That makes
our framework directly applicable to those models. Specifically,  our
operator spin foam framework can be translated into the language of spin foams and partition functions. We discuss the examples: BF spin foam model, the BC model, and the model obtained by application of our framework to the EPRL intertwiners.   
\end{abstract}

\pacs{
  {04.60.Pp}
  }
\maketitle

\def\be{\begin{equation}}
\def\ee{\end{equation}}
\def\ba{\begin{eqnarray}}
\def\ea{\end{eqnarray}}
\def\lp{{\ell}_{\rm Pl}}
\def\g{\gamma}
\section{Introduction}
The successful application of the 3d BF spin-foam theory to 3d quantum gravity
(see \cite{Baezintro,PR} and references therein) produced  and still produces
activity in the 4d spin-foam approaches to the 4d quantum
gravity\cite{Baezintro}-\cite{ThiemannSpinFoams}. After the decade of the Barrett
Crane model \cite{BC}, a breakthrough has come with the new models: the
Engle-Pereira-Rovelli-Livine model \cite{EPRL,flipped} and  the Freidel-Krasnov
model \cite{FK}. For the first time, the existence of a relation between the 4d
spin-foam theory on the one hand, and the kinematics of the 3+1 loop quantum
gravity \cite{Ashtekarbook,Thiemannbook,Rovellibook,Marev,AshLewrev} has become
plausible. The theory accommodates all the states of LQG labelled by graphs
embedded in an underlying 3-manifold \cite{SFLQG} although seems not to be
sensitive on linking and knotting   \cite{Benjamin}.

The spin networks and spin foams featuring in the spin foam models may be thought of as just  combinatorial tools used to extract numbers. However, they also admit their  own structure that deserves understanding. The spin networks emerge in loop quantum gravity  as invariant elements of the tensor products of representations.  Consistently, the spin foams arise as cobordisms between the spin networks, and hence should be described in terms of operators mapping the invariants into invariants.

The goal of this paper is introducing a systematic approach to spin foams.
We introduce  operator spin foams, that is foams labelled by group representations
and operators, as the main tool. An equivalence relation we define in the set of the operator spin foams allows to split the faces and the edges of the foams. The consistency with that relation requires introduction of the (familiar for the BF theory) face amplitude. The operator spin foam models are defined quite generally. Imposing a maximal symmetry leads to a family we call natural operator spin foam models.   This symmetry, combined with demanding consistency with splitting the edges, determines  a complete characterization of a general natural model.
It can be obtained by applying arbitrary (quantum) constraints on an arbitrary BF spin foam model. Remarkably, imposing suitable constraints on Spin(4) BF spin foam model is exactly the way we tend to view 4d quantum gravity, starting with the BC model and continuing with the EPRL or FK models. That makes
our framework directly applicable to those models. Specifically,  our
operator spin foam framework can be translated into the language of spin foams and partition functions. Then, it offers a definition of the partition function
for the EPRL model as well. The result is that of \cite{cEPRL}, rather than the one defined in the original EPRL paper \cite{EPRL}. The choice of the EPRL intertwiners and the vertex amplitude is the same in both approaches. The ambiguity is in glueing
the vertices. Of course we do not mean to insist that the proposal of \cite{cEPRL}
that also follows from the current paper is better than the original EPRL one. We just find a set of natural properties that lead to the former proposal, and
the bottom line is, that the latter proposal is necessarily inconsistent with one of the conditions we spell out.

\section{Operator spin foam}
\subsection{Definition}\label{subsecOSFD}
Let $\kappa$ be a locally linear, oriented
2-complex with boundary $\partial \kappa$
\cite{Baezintro,SFLQG}  and let $G$ be a compact Lie group.
Denote by $\kappa^{(0)}$ the set of vertices (the 0-cells), by $\kappa^{(1)}$ the set of edges (1-cells)
and by $\kappa^{(2)}$ the set of faces (2-cells) of the complex $\kappa$.
For simplicity of the presentation, we will be assuming throughout this paper
that every face of $\kappa$ is topologically a disc.\footnote{That is no point of a face is glued to another point of a same face; below we introduce an equivalence relation which allows to split/glue faces and edges. It will be obvious how to use those moves to relax this assumption.}
Every edge $e\in \kappa^{(1)}$ is contained in at least one face. If $e$ is contained
in exactly one face, we call it boundary edge. Otherwise $e$ is an internal edge.
If a vertex $v\in\kappa^{(0)}$ is contained in a boundary edge, we call it boundary vertex. Otherwise $v$ is internal. We will be denoting the set of internal edges/vertices  by ${\rm int}\kappa^{(1)}$ / ${\rm int}\kappa^{(0)}$.

The 1-complex set by the  boundary edges and boundary vertices is denoted by
$\partial \kappa$  and called the boundary of $\kappa$.

\begin{figure}[hbt!]
  \centering
\subfloat[Colouring of
faces]{\label{fig:faces}\includegraphics[width=0.45\textwidth]{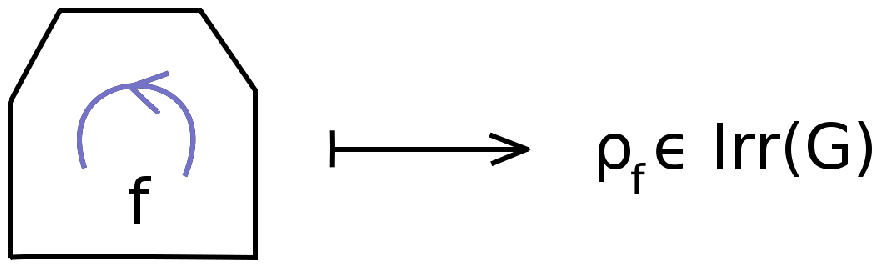} }
\subfloat[Colouring of
edges]{\label{fig:edges}\includegraphics[width=0.35\textwidth]{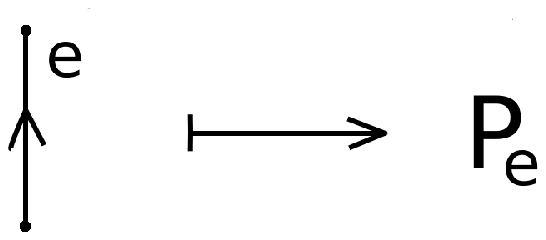} }
  \caption{Operator form of Spin-Foam}
  \label{fig:operator}
\end{figure}

An operator-spin-foam we define in this paper is a triple
$(\kappa,\rho,P)$, where  $\rho$ and $P$ are colorings by representations and, respectively, operators defined below. The first one, $\rho$  is  familiar from  spin-foam theories, namely
\begin{itemize}
\item $\rho$ is a coloring of the faces with irreducible
representations of $G$ (fig. \ref{fig:faces}),
\ba\rho:\kappa^{(2)}&\rightarrow {\rm Irr}(G),\\
                    f&\mapsto \rho_f. \ea
\end{itemize}
The coloring $\rho$ can be used to assign Hilbert spaces to the faces and the edges of $\kappa$. To every face $f$, there is assigned a Hilbert space ${\cal H}_f$
\be f\mapsto {\cal H}_f\ee
on which the representation  $\rho_f$ acts.
To every edge $e$ there is assigned a Hilbert space $H_e$ defined by
the Hilbert spaces of the faces containing $e$,
\be{\cal
H}_{e}\ =\ \bigotimes_{f\textrm{ incoming to }e} {\cal H}^*_f \otimes
\bigotimes_{f'\textrm{ outgoing from }e} {\cal H}_{f'}\label{He}\ee
where, a face is  called incoming to (outgoing from) an edge $e$
if its orientation agrees with (is opposite to) that of $e$, and by $\Hil_f^*$ we
denote the algebraic dual. Given a representation $\Hil$ of $G$ (irreducible), the subspace of invariant elements is denoted by ${\rm Inv}\Hil$.

Having in mind those Hilbert spaces we introduce the
operator labelling:
\begin{itemize}
\item $P$ is a colouring of the internal edges with operators (fig. \ref{fig:edges})
\begin{eqnarray} {\rm int} \kappa^{(1)}\ni e\ &\mapsto P_e\label{Pe}\\
P_e:{\rm Inv}{\cal H}_{e}\ &\rightarrow\ {\rm Inv}\Hil_e. \end{eqnarray}
\end{itemize}

\begin{figure}[hbt!]
  \centering
 \includegraphics[width=\textwidth]{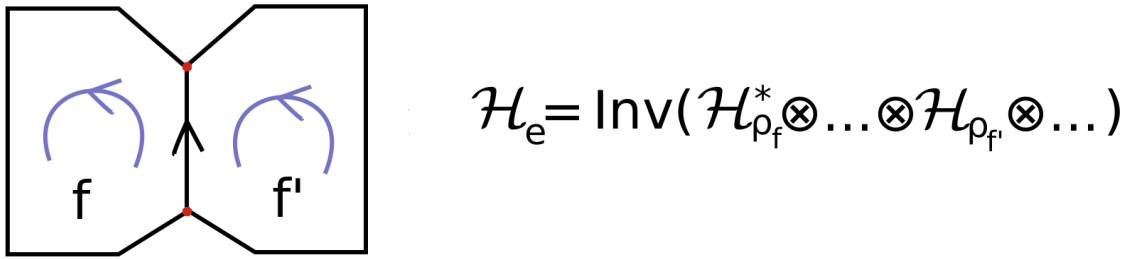}
\caption{The edge Hilbert space ${\cal H}_e$}
\end{figure}

\subsection{Equivalence relation}\label{subsecER}
In the space of operator-spin-foams we introduce an equivalence relation that reflects the equivalence relation in the space of the spin networks. These relations
allow to subdivide edges and faces and also change their orientation.
In the four following paragraphs we describe that equivalence relation of operator-spin-foams in detail.

\subsubsection{Edge reorientation}\label{subsubsecERER}
\begin{figure}[hbt!]
  \centering  
\includegraphics[width=0.8\textwidth]{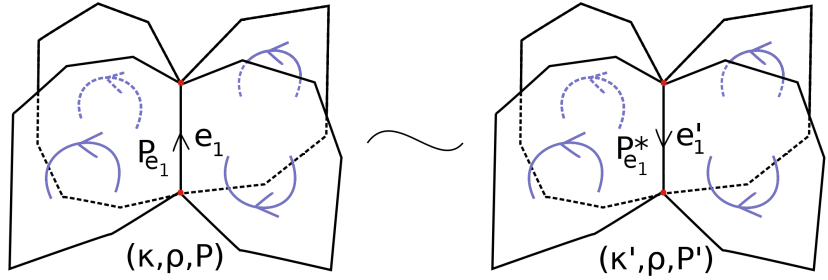}
\caption{Invariance under the face subdivision}
\end{figure}
Given an operator-spin-foam $(\kappa,\rho,P)$, let us switch the orientation of its edge $e_1$,
\be e'_1\ =\ e_1^{-1},\ee
and leave all the other orientations unchanged. Denote the resulting 2-complex by $\kappa'$. To define an operator-spin-foam $(\kappa',\rho',P')$ which is equivalent to $(\kappa,\rho,P)$, suppose first that the edge $e_1$ is internal and
\begin{itemize}
\item leave the labelling $\rho$, namely
\be\rho'=\rho.\ee
\end{itemize}
 Now,  $\rho'$ determines the Hilbert space $\Hil_{e'_1}$ to be
\be \Hil_{e'_1}\ =\ \Hil_{e_1}^* \ee
where the algebraic dualization $*$ is applied to each factor on in the right hand side of (\ref{He}). The natural choice for ${P'}_{e'_1}$ is
\begin{itemize}
\item for the reoriented edge $e'_1={e_1}^{-1}$,
\be {P'}_{e'_1}\ =\ P_{e_1}^*, \ee
\item whereas for the remaining edges of $\kappa'$ we leave
\be {P'}_e\ =\ P_e .\ee
\end{itemize}
The operator spin foams $(\kappa,\rho,P)$ and $(\kappa',\rho,P')$ are equivalent,
\be (\kappa,\rho,P) \ \equiv\ (\kappa',\rho,P'). \ee

The remaining case when the reoriented edge $e_1$ is internal is yet simpler:
both labellings $\rho$ and $P$ are defined on the faces/edges unaffected
by the reorientation of $e_1$;  we just leave them unchanged, that is
we set $\rho'=\rho$ and $P'=P$.

\subsubsection{Face reorientation}\label{subsubsectERFR}
\begin{figure}[hbt!]
  \centering  
\includegraphics[width=0.8\textwidth]{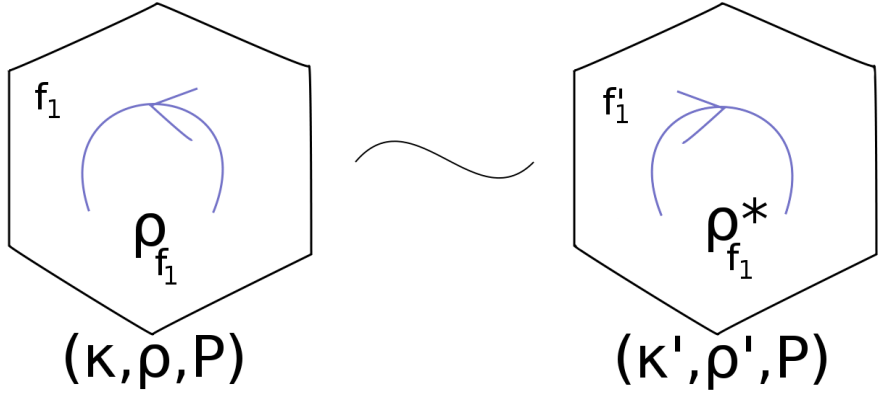}
\caption{Invariance under the face subdivision}
\end{figure}
Given an operator-spin-foam $(\kappa,\rho,P)$, let us switch the orientation of its face $f_1$ and denote the reoriented face  $f'_1$. Denote the resulting 2-complex by $\kappa'$. To define an operator-spin-foam $(\kappa',\rho',P')$  equivalent to $(\kappa,\rho,P)$, we modify  the labelling $\rho$ in the following way:
\begin{itemize}
\item for the reoriented face $f'_1$ we take the dual representation,
\be\rho'_{f'_1}\ =\ \rho_{f_1}^*,\ee
\item  for the remaining faces, the labelling $\rho'$ coincides with $\rho$,
\be \rho'_f\ =\ \rho_f, \ {\rm for}\ f\not=f'_1.\ee
\end{itemize}
At each edge $e$, the labelling $\rho'$ defined the same Hilbert space $\Hil_{e}$
as $\rho$ in $(\kappa,\rho,P)$. Therefore, the following definition of $P'$
is possible,
\begin{itemize}
\item For a labelling ${P'}$ the choice is
\be {P'}\ =\  P. \ee
\end{itemize}
Again, we will consider $(\kappa',\rho',P)$ and $(\kappa,\rho,P)$ equivalent,
\be (\kappa,\rho,P) \ \equiv\ (\kappa',\rho,P'). \ee

\subsubsection{Face splitting}\label{subsubsectERFS}
\begin{figure}[hbt!]
  \centering
\includegraphics[width=0.8\textwidth]{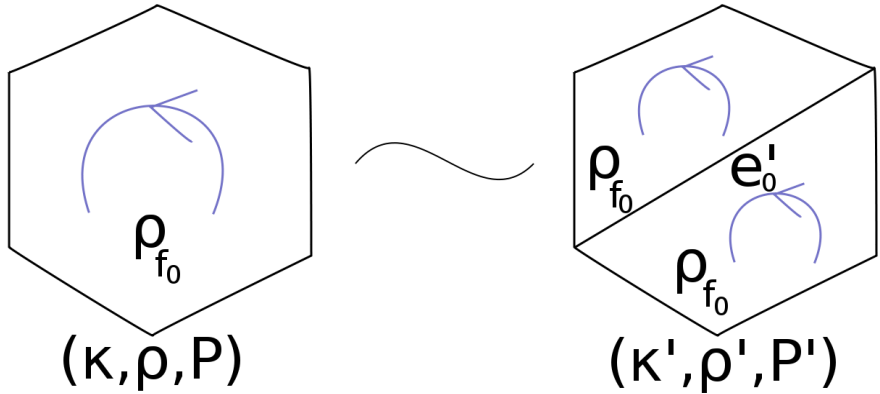}
\caption{Invariance under face subdivision}
\label{face_division}
\end{figure}
Consider an operator spin-foam $(\kappa,\rho,P)$.
Split one of its faces, $f_0$ say, into $f'_1$ and $f'_2$ such that a resulting
new edge $e'_0$ (oriented arbitrarily) contained in $f'_1$ and in $f'_2$ connects two vertices belonging to $\kappa^{(0)}$. Choose an orientation of the new faces to be the one induced by $f_0$. The resulting new 2-cell complex $\kappa'$ is obtained by replacing the face $f_0$ by the pair of faces $f'_1$ and $f'_2$ and by adding the edge $e'_0$. Define a labelling $\rho'$ on $\kappa'$ in the following way
\begin{itemize}
\item $\rho'$ coincides with $\rho$ on the unsplitted faces,
\begin{equation} \rho'_{f'}\ = \rho_{f'},\ {\rm if}\ f'\not= f'_1,f'_2\ee
\item and $\rho'$ agrees with $\rho$ on the faces $f'_1,f'_2$ resulting from the
splitting
\be \rho'_{f'}\ =\ \rho_{f_0},\ {\rm if}\ f'=f'_1,f'_2                        \end{equation}
\end{itemize}
For the edge $e'_0$,  the corresponding Hilbert space is one dimensional by Schur's Lemma,
\be \Hil_{e'_0}\ =\ {\rm Inv}\left(\Hil_{f_0}\otimes \Hil_{f_0}^*\right)\equiv\ \mathbb{C}. \ee
Define a labelling $P'$ of the edges of $\kappa'$
\begin{itemize}
\item to be the identity on the new edge $e'_0$ resulting from the splitting,
\be  P'_{e'}\ =\  {\rm id},\ {\rm if}\ e' = e'_0\ee
\item and to coincide with $P$ on the old edges
\be P'_{e'}\ =\ P_{e'},\ {\rm if}\ e'\not= e'_0\,. \ee
\end{itemize}
The resulting operator spin foam is equivalent to $(\kappa,\rho,P)$,
\be (\kappa,\rho,P) \ \equiv\ (\kappa',\rho,P'). \ee

\subsubsection{Edge splitting}\label{subsubsecERES}
\begin{figure}[hbt!]
  \centering
\includegraphics[width=0.8\textwidth]{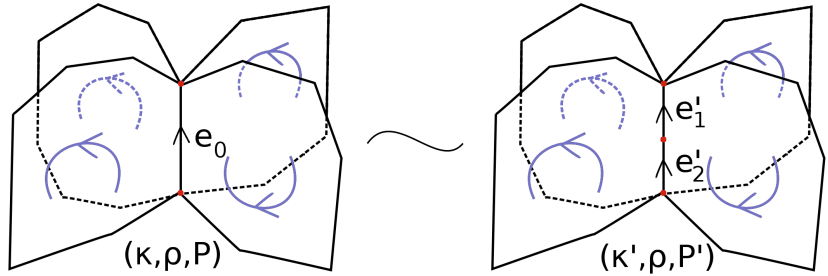}
\caption{Invariance under the edge subdivision}
\label{edge_division}
\end{figure}
In an operator spin foam $(\kappa,\rho,P)$ split an edge $e_0$ into $e'_1$ and $e'_2$
\be e_0\ =\ e'_2\circ e'_1 \ee
whose orientations are induced by $e_0$. Denote the resulting 2-complex by $\kappa'$.
An operator spin foam $(\kappa',\rho'.P')$ defined on $\kappa'$ is equivalent to
 $(\kappa,\rho,P)$,
 \be (\kappa,\rho,P) \ \equiv\ (\kappa',\rho',P')\,, \ee
  whenever the following conditions are satisfied by $\rho'$ and $P'$:
\begin{itemize}
\item   $\rho$ is unchanged,
\be \rho'\ =\ \rho,\ee
\item $P'$ coincides with $P$ on the edges $e'\not= e'_1,e'_2$,
\item $P'_{e'_1}$ and $P'_{e'_2}$ satisfy the following constraint
\be P'_{e'_2}\circ P'_{e'1}\ =\ P_{e_0}, \ee
provided the edge $e_0$ is internal.
\end{itemize}

\subsubsection{Rescaling of the operators}
Every  operator spin foam $(\kappa,\rho,P)$ is equivalent to any operator spin foam
$(\kappa,\rho,P')$ defined by rescaling, for every internal edge $e$, 
\be P'_e\ =\ a_e P_e,\ \ \ \ \  a_e\in \mathbb{C},\ee
such that
\be \prod_ea_e\ =\ 1.\ee

\subsubsection{Adding a face labelled by the trivial representation}
Our definition of the operator spin foams does not exclude the trivial representation
from the set of labels assigned to the faces. Every spin foam $(\kappa,\rho,P)$
will be considered equivalent to a spin foam $(\kappa',\rho',P')$ obtained
by adding a face $f'_1$ and labelling it by the trivial representation $\rho_0$. That is,
\begin{eqnarray} \rho'(f')\ =\begin{cases} \rho(f'), &{\rm if}\ f'\in \kappa^{(2)}\\
                              \rho_0,   &{\rm if}\ f'=f'_1.\end{cases}      
\end{eqnarray} 
All the internal edges $e$ and the corresponding Hilbert spaces $\Hil_e$
coincide, and $P'$ is defined to be,
\be P'=P.\ee 

\subsection{Glueing the operator spin foams}\label{subsecGOSF}
In the space of the 2-complexes considered in this paper there is the obvious operation of glueing. It admits a natural extension to an operation of glueing the operator spin foams, which we describe in the following, for the sake of completeness.
Two  oriented, locally linear 2-cell complexes $\kappa$ and $\kappa'$ can be glued along a connected component $\gamma$ of the boundary $\partial \kappa$ and a connected component $\gamma'$ of $\partial \kappa$, provided $\gamma$ and $\gamma'$ are isomorphic closed 1-cell complexes (unoriented graphs), and the orientations of the glued faces and, respectively,  their sites match. If $\phi:\gamma\rightarrow\gamma'$ is an isomorphism, then the glueing amounts to glueing along each link $e$ of $\gamma$ a face $f_e$ of $\kappa$ containing $e$, with the face $f'_{\phi(e)}$ of $\kappa'$ containing the link $\phi(e)$ of $\gamma'$.
In what follows we will assume that the map
\be \gamma\ni e\mapsto f_e, \ \ \gamma'\ni e'\mapsto f'_{e'} \ee
is 1-1 (each $e$ has its own $f_e$). This can be always achieved by dividing
the faces and edges. The resulting face
$f_e\# f'_{\phi(e)}$ can be oriented either according to the orientation of $f_e$ or to the orientation of $f'_{\phi(e)}$; coinciding of the  two orientations is the matching relation we have mentioned above. A similar matching condition applies
to the oriented sides of the faces $f_e$ and $f'_{\phi(e)}$.  Repeating that glueing
for every link $e$ of $\gamma$, we complete the glueing of $\kappa$ and $\kappa'$
along $\gamma$. The result can be denoted by $\kappa\#\kappa'$ and it depends
on the graphs $\gamma$, $\gamma'$ and the isomorphism $\phi$. If the 2-complexes above were endowed with the structures of the operator spin foams $(\kappa,\rho,P)$, and respectively, $(\kappa', \rho',P')$, the operator spin foams can be glued into an operator spin foam $(\kappa\#\kappa',\rho\#\rho',P\# P')$ provided the representations
agree on the boundary, and the glueing condition is
\be \rho'_{f'_{\phi(e)}}\ =\ \rho_{f_e} \ee
for every pair $e$ and $\phi(e)$ of the identified edges.
\begin{itemize}
\item For every of the boundary edges $e$, due to the glueing condition we can
set
\be (\rho\#\rho')_{f_e\# f'_{\phi(e)}}\ =\ \rho_{f_e}\ =\ \rho'_{\phi(e)}. \ee
\item For the remaining faces we use either $\rho$ or, respectively, $\rho'$
\be (\rho\#\rho')_{f''}\ = \begin{cases}\rho_{f''},\ &{\rm if}\ f''\in\kappa^{(2)},\\
\rho'_{f''}\ &{\rm if}\ f''\in\kappa'^{(2)},\end{cases}.\ee
\end{itemize}
For the operator part $P\# P'$,  the glueing consists in
\begin{itemize}
\item taking the composition of the operators for every
pair $(\tilde{e},\tilde{e'})$ of sides of the faces $f_e$, and respectively, $f'_{\phi(e)}$ that are glued
into a side of the face $f_e\# f'_{\phi(e)}$,  that is either
\be (P\# P')_{\tilde{e}\circ\tilde{e'}}\ =\ P_{\tilde{e}}\circ P_{\tilde{e'}}\ee
or
\be (P\# P')_{\tilde{e'}\circ\tilde{e}}\ =\ P_{\tilde{e'}}\circ P_{\tilde{e}}\ee
depending on the orientations.
\item For each of the remaining edges of $\kappa\#\kappa'$ we leave the corresponding
operator of either $\kappa$ or $\kappa'$,
\begin{equation} (P\# P')_{e''}\ =\ \begin{cases}P_{e''}, &{\rm if}\ e''\in{\rm int}\kappa\\
                                    P'_{e''}, &{\rm if}\ e''\in{\rm int}\kappa'.\end{cases}\end{equation}
\end{itemize}

\section{Spin foam operator}
\subsection{2-edge contraction}\label{subsec2EC}
\begin{figure}[ht!]
  \centering
    \includegraphics[width=0.85\textwidth]{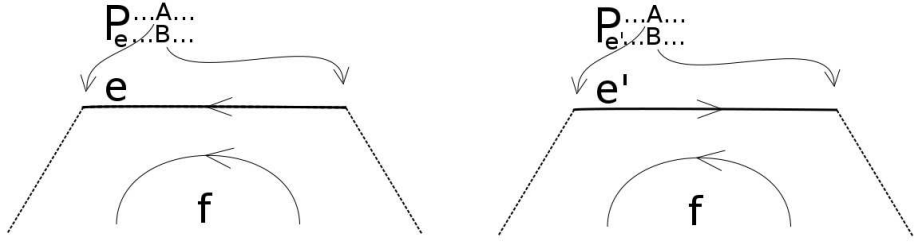}
\caption{The rule of assigning an index of $P_e$ to a corner $v$ of a face $f$: given an edge $e$ contained in
a face $f$ of an operator-spin-foam $(\kappa,\rho,P)$,
in the operator  $P_e$, the indices corresponding to the Hilbert space
${\cal H}_f$ of the representation $\rho_f$ are assigned to the end points of $e$
such that the lower / upper index is assigned to the point that is  the beginning / end point of $e$ if the orientation of $e$ is the same as that of $f$, and to the end / beginning  point of $e$ if the orientation of $e$ is opposite.   The oriented arc only
marks the orientation of the polygonal face $f$.}
{\label{indices}}
\end{figure}

\begin{figure}[hbt!]
  \centering
\includegraphics[width=0.4\textwidth]{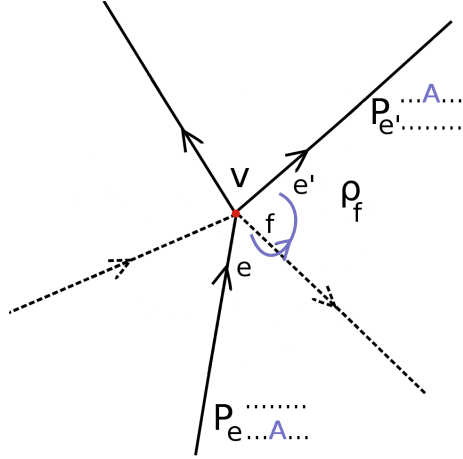}
 \caption{{2-edge contraction of indices: The edges $e$ and $e'$ are connected
 by the face $f$. The blue indices $A$ of $P_{e}$ and, respectively  $P_{e'}$
 correspond to the Hilbert space $\Hil_f$ and get contracted by ${\rm Tr}_{v,f}$.}}
\label{contraction}
\end{figure}

 Wherever two internal edges of a spin-foam $(\kappa,\rho,P)$ meet, the geometry of a spin-foam defines a natural contraction between the corresponding operators.
The easiest way to introduce it is to use the (abstract) index notation. It is as follows: given
\be w\in {\rm Inv}\left(\bigotimes_{f\textrm{ incoming to }e} {\cal H}^*_f
\otimes \bigotimes_{f'\textrm{ outgoing from }e} {\cal H}_{f'}\right)\ee
we denote it in the index notation as
\be w \ =\ w_{A...}{}^{A'...}\ee
where the lower/upper indices  correspond to the spaces $\Hil_f^*$ / $\Hil_{f'}$.
The action of the operator $P_e$ reads
\be (P_e w)_{A...}{}^{A'...}\ =\ {P_e}_{A...B'...}^{A'...B...}w_{B...}{}^{B'...}.\label{Pw}\ee
Moreover, the vector $w_{A...}{}^{A'...}$ is associated to the beginning of the given edge $e$, whereas the vector  $(P_e w)_{A...}{}^{A'...}$ lives at the end of $e$.
In this sense, the indices $B,B'$ of ${P_e}_{A...B'...}^{A'...B...}$ are associated
with the beginning point of $e$, whereas the indices $A,A'$ of ${P_e}_{A...B'...}^{A'...B...}$ with the end point of $e$. Therefore, for every edge $e$, in the operator $P_e$, for each face $f$ containing $e$, there are two indices, an
upper and a lower one corresponding to the Hilbert space ${\cal H}_f$. The indices are associated with the ends
of the edge $e$, according to the rule introduced above and   presented in FIG. \ref{indices}. In the figure, we did not indicate an orientation of the edge $e$, because it does not affect the position of the indices of $P_e$. The position of the indices depends only on the orientation of a given face $f$, and for every face,
every edge contained in it and every labelling it is exactly as in FIG. \ref{indices}.     
\\
Now, for every pair of edges $e$ and $e'$ which belong to the same face $f$
and share a vertex $v$, if the index of $P_e$ corresponding to $f$ and $v$
is upper / lower, then the  index of  $P_{e'}$
corresponding to $f$ and $v$ is lower / upper, respectively. In this way, there is defined the natural contraction Tr$_{v,f}$ at $v$
(FIG. \ref{contraction}).

\subsection{Contracted operator spin foam}
The contraction at the vertices of the complex defines the  contracted
operator-spin-foam:
{
\begin{equation}\label{tr}
{\rm Tr}{(\kappa,\rho,P)}\ :=\
\prod_{v,f}{\rm
Tr}_{v,f}\left(\bigotimes_{e\in{\rm Int}\kappa^{(1)}}P_e\right)   \end{equation}
Given an edge $e$, one of its ends $v$ and a face $f$ containing $e$,
the corresponding index in $P_e$ is contracted, provided there is another
internal (that is contained in at least two different faces) edge $e'$ contained in $f$ and intersecting the point $v$. Otherwise, the index stays uncontracted.
As a consequence, the contracted operator ${\rm Tr}{(\kappa,\rho,P)}$ is indeed an operator.
Identifying each operator $P_e:\Hil_e\rightarrow\Hil_e$ with an element of $\in
\Hil_e\otimes\Hil_e^*$, the contracted spin-foam  ${\rm Tr}{(\kappa,\rho,P)}$
is identified with an element of the Hilbert space
\be \Hil_{\rm \partial \kappa}\ =\ \bigotimes_{e\ incoming\ to\ \partial \kappa }\Hil_e\otimes \bigotimes_{e'\ outgoing\ from\ \partial\kappa}\Hil_{e'}^*. \ee }

\subsection{Spin foam operator}
\subsubsection{Contraction and equivalence}
Any splitting $\Hil_{\partial\kappa}=\Hil_{\rm fin}\otimes \Hil_{\rm in}^*$
makes the contracted operator spin foam  ${\rm Tr}{(\kappa,\rho,P)}$ an
operator $\Hil_{\rm in}\rightarrow \Hil_{\rm fin}$.

There is a catch, however. The expression  (\ref{tr}) should respect the operator spin foam equivalence relation, i.e. Tr should only depend on equivalence classes of operator spin foams. Given an operator spin foam $(\kappa,\rho,P)$
suppose an equivalent operator spin foam $(\kappa',\rho',P')$ is obtained
from $(\kappa,\rho,P)$ by either the reorientation or by the edge splitting
as in Section \ref{subsubsecERER}, \ref{subsubsectERFR} or \ref{subsubsecERES} .
Then
\be {\rm Tr}(\kappa',\rho',P)\ =\ {\rm Tr}(\kappa,\rho,P). \ee
However, if an operator spin foam $(\kappa',\rho',P')$ equivalent to
$(\kappa,\rho,P)$ is obtained by splitting a face $f_0$ of $\kappa$ and defining
$\rho'$ and $P'$ as in Section \ref{subsubsectERFS}, then the equivalence is
not preserved by
the trace. In that case,  the Hilbert space
$${\rm Inv}\left(\Hil_{f'_1}\otimes \Hil_{f'_2}^*\right)={\rm Inv}\left(\Hil_{f_0}\otimes\Hil_{f_0}^*\right)$$
is spanned by the element, in the index notation,  $\delta^a_b$, and    the operator $P'_{e'_0}=1$ reads
\be P'_{e'_0}{}^{ab'}_{a'b}\ =\ \frac{1}{d_{f_0}}\delta^a_b\delta^{b'}_{a'}. \ee
It is easy to verify that
\be {\rm Tr}(\kappa',\rho',P)\ =\ \frac{1}{d_{f_0}}{\rm Tr}(\kappa,\rho,P) \ee
where
\be d_{f_0}\ =\ {\rm dim}\Hil_{f_0}.\ee
Hence the equivalence relation is not preserved.

\subsubsection{Face amplitude restores the equivalence}
Introducing suitable face amplitude makes the contraction ${\rm Tr}$ of
operator spin foam exactly compatible with the equivalence relation.
Consider a spin foam operator defined by a formula (tilde will be removed
when we establish the final form of the operator)
\be \tilde{\cal Z}_{(\kappa,\rho,P)}\ =\ \left(\prod_{f\in\kappa^{(1)}}A_f\right){\rm Tr}(\kappa,\rho,P)\label{tildeZ} \ee
where
$$f\mapsto A_f$$
is an unknown function, a face amplitude. Then,
a unique solution for $f\mapsto A_f$ such that for every operator spin foam
$(\kappa,\rho,P)$ and every equivalent operator spin foam $(\kappa',\rho',P')$
\be \tilde{\cal Z}_{(\kappa,\rho,P)}\ =\
\tilde{\cal Z}_{(\kappa',\rho',P')},\ee
is
\be A_f\ =\ {\rm dim}\Hil_f\,. \label{Af}\ee

\subsubsection{Boundary amplitude restores the compatibility with the glueing}
\begin{figure}[ht!]
  \centering
\includegraphics[width=0.4\textwidth]{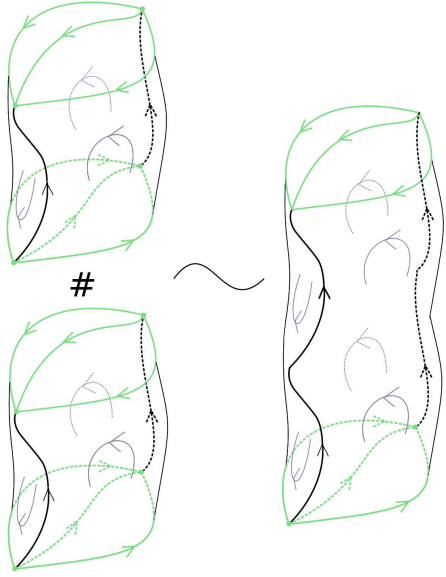}
\caption{Compatibility with the glueing of the operator spin foams}
\label{composition}
\end{figure}
The introduction of the face amplitude destroys  the compatibility
with the glueing of the operator spin foams. Consider two operator spin foams
$(\kappa,\rho,P)$ and $(\kappa',\rho',P')$, and their  composition
$(\kappa,\rho,P)\#(\kappa',\rho',P')$ glued along a graph $\gamma$.
The operator spin foam contraction induces the contraction of the operators
$\tilde{\cal Z}(\kappa,\rho,P)$ and  $\tilde{\cal Z}(\kappa',\rho',P')$,
let us denote it by  ${\rm Tr}_\gamma$. The result is
\be {\rm Tr}_\gamma\left( \tilde{\cal Z}(\kappa,\rho,P)\otimes\tilde{\cal Z}(\kappa',\rho',P') \right)\ =\ \prod_{e\in\gamma}A(f_e)\tilde{\cal Z}(\kappa\#\kappa',\rho\#\rho',P\# P'). \ee
To restore the compatibility of $\tilde{\cal Z}$ with glueing the operator spin foams
we finally define the spin foam operator to be
\be {\cal Z}(\kappa,\rho,P)\ :=\ \prod_{e\in(\partial\kappa)^{(1)}}\frac{1}{\sqrt{A_{f_e}}}\tilde{\cal Z}(\kappa,\rho,P), \label{Z}\ee
where $f_e$ is the face of $\kappa$ containing $e$ (and we are assuming
that $e\not=e'\Rightarrow f_e\not=f_{e'}$ that can be always achieved by splitting faces and edges.). Now we have
\be {\rm Tr}_\gamma\left( {\cal Z}(\kappa,\rho,P)\otimes{\cal Z}(\kappa',\rho',P') \right)\ =\ {\cal Z}(\kappa\#\kappa',\rho\#\rho',P\# P'). \ee

\subsection{Relation with the spin foams and state sums}
\subsubsection{The spin foams}
The operator spin foam formalism seem to differ from the usual formulation of spin foam amplitudes, in that there are projection operators assigned to edges instead of intertwiners. However, the projection operators $P_e$ can be interpreted as the result of spin foam amplitudes where the sum over the intertwiners has already been carried out, i.e. we decompose each $P_e$,
\be P_e\ =\sum_{\iota_e\in{\cal B}_e}\sum_{\iota_e'\in{\cal B}^\dagger_{e'}}\ P_{\iota_e}^{\iota_e'}\, \iota_{e}\otimes\iota'_{e}\label{Peii}\ee
in any basis,
\be {\cal B}_e\subset \Hil_e,\ee
and the conjugate basis
\be{\cal B}_e^\dagger=\{\iota_e^\dagger:\iota_e\in{\cal B}_e\}\subset\Hil_e^*,\label{B'}\ee
where $\Hil\ni v\mapsto v^\dagger \in \Hil^*$ is the canonical antilinear map
(denoted by $|v\rangle\mapsto\langle v|$ in the Dirac notation).

After the  substitution of the right hand side of (\ref{Peii}) for $P_e$, the tensor product $\bigotimes_{e}P_e$ becomes a linear combination of the tensor products
\be \bigotimes_{e}\iota_{e}\otimes\iota'_{e},\ee
in which to each internal edge $e$ there is assigned a (tensor product of a) pair of the intertwiners $\iota_{e}\otimes\iota'_{e}$, where $\iota_e\in{\cal B}_e$
and $\iota'_e\in{\cal B}_e^\dagger$ are independent of each other. In fact, from the point of view of the contractions we use, $\iota'_e$
is assigned to the beginning point of $e$ whereas $\iota_e$ is assigned to the end point of $e$. That is the generalised case of a spin foam that was derived in \cite{cEPRL}.

\subsubsection{The vertex amplitude}
Given a vertex $v$, the application of the constructions ${\rm Tr}_{vf}$  (see
Section \ref{subsec2EC}) for all the  faces $f$ which intersect $v$, namely
\be\prod_{f\ :\ f\ni v}{\rm Tr}_{vf}\left(\bigotimes_{e}\iota_{e}\otimes\iota'_{e}\right),\ee
produces a $\mathbb{C}$ number factor
\be A_v\ =\ \prod_{f\ :\ f\ni v}{\rm Tr}_{vf}\left( \bigotimes_{e\ {\rm incoming}}\iota_e\otimes\bigotimes_{e'\ {\rm outgoing}}\iota'_{e'}\right),\label{Av} \ee
where  $e/e'$ ranges the set of edges that end/begin at $v$ and each  $f$ connects a pair of the edges (either two unprimed, or two primed, or one primed and one unprimed). The factor $A_v$ is known in the spin foam literature as the  vertex amplitude.

\subsubsection{The state sums}
\begin{figure}[ht!]
  \centering
\includegraphics[width=0.65\textwidth]{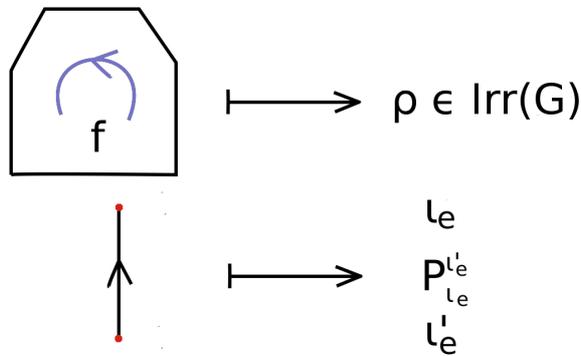}
\caption{The operator approach is equivalent to approach in which we
assign an irreducible representations of group $G$ to each face of the
2-complex and a pair of intertwiners $\iota_e$,$\iota_e'$ together with the
complex number $P_{\iota_e}^{\iota_e'}$ to each internal edge.}
\end{figure}
Finally,  the substitution of the right hand side of (\ref{Peii}) into the spin foam
operator ${\cal Z}(\kappa,\rho,P)$ definition (\ref{tr},\ref{tildeZ},\ref{Af},\ref{Z}) gives the following sum with respect to all the labellings of the internal edges $e\in {\rm int}\kappa$,
\be\iota : e\mapsto \iota_e\otimes \iota'_e\in {\cal B}_e\otimes{\cal B}_e^\dagger,\ee
namely
\be
{\cal Z}(\kappa,\rho,P)\ = \sum_{\iota}\prod_e P^{\iota'_e}_{\iota_e}\prod_{f}d_f\prod_{v}A_v
\prod_{\tilde{l}} \frac{1}{\sqrt{d_{f_{\tilde{l}}}}}
\bigotimes_{{\tilde e}}\iota_{{\tilde e}}\otimes \bigotimes_{{\tilde e}'}\iota'_{{\tilde e}'}\label{statesum}
\ee
where $f$ runs through the set of faces  and $d_f$ is the dimension of $\rho_f$, $v$ ranges the set of the internal vertices, $l$ ranges the set of the boundary edges (links) and $f_{l}$ is the face containing ${l}$, and ${\tilde e}$/${\tilde e}'$ ranges the set of edges which intersect $\partial \kappa$ at the end/beginning point.
Finally, the familiar partition function emerges in the formula
\begin{equation}
{\cal Z}(\kappa,\rho,P)\ =\ \sum_{\iota_{\partial{\kappa}}}Z(\kappa,\rho,\partial\iota)
\bigotimes_{e}\iota_e\otimes \bigotimes_{e'}\iota'_{e'}, \end{equation}
where $e$ and $e'$ rages the same set of edges as in the previous equality,
and $\partial\iota$

\section{Operator spin foam models}
\subsection{Definition, natural models}
\subsubsection{Definition}
A $G$  operator spin foam model, where $g$ is a compact group, can be defined as an assignment of an operator spin foam $(\kappa,\rho,P)$ to each locally linear 2-complex $\kappa$ endowed with a labelling $\rho$ of the faces of $\kappa$ with
the irreducible representations of  $G$ (see  Sec \ref{subsecOSFD}),
\be (\kappa,\rho)\ \mapsto \ (\kappa,\rho,P),\label{P} \ee
which preserves the equivalence relation of Sec \ref{subsecER} and is consistent with the glueing operation of Sec \ref{subsecGOSF}.

\subsubsection{Natural operator spin foam models}
We will consider below  a class of natural  operator spin foam models, that is models such that, briefly speaking,
\begin{itemize}
\item the assignment $e\mapsto P_e$ depends only on the unordered sequence of labels $\rho_f$ such that $e\subset f$
\end{itemize}
and is independent of the other parts of a given 2-complex $\kappa$ - see below for a technical definition. We will be also assuming that the assignment $P$ is self-adjoint, that is
\begin{itemize}
\item for every edge internal $e\in {\rm int}\kappa^{(1)}$
\be P_e^\dagger\ =\ P_e,\ee
\end{itemize}
(of course $P_e$ is defined only for the internal edges).

Technically, the first assumption means, that for every unordered sequence $R$ of irreducible  representations of the group $G$, we fix an operator
\be P_R\ :\ {\rm Inv}\bigotimes_{\rho\in R}\Hil_\rho\ \rightarrow\ {\rm Inv}\bigotimes_{\rho\in R}\Hil_\rho.\label{PR}\ee
Moreover, given two sequences $R$ and $R'$ which coincide modulo a representation
$\rho_1\in R$ and $\rho'_1\in R'$ (that is after removing $\rho_1$ from $R$ and $\rho'_1$ from $R'$, the remaining unordered sequences coincide) and given an intertwining operator
\be \iota\circ \rho_1\ =\ \rho'_1\circ\iota \ee I
the corresponding operators are also intertwined
\be \left(\iota\otimes \bigotimes_{\rho\in R, \rho\not=\rho_1}{\rm id}_\rho\right)\,\circ\, P_R\ =\
P_{R'}\circ \left(\iota\,\otimes\,\bigotimes_{\rho'\in R', \rho'\not=\rho'_1}{\rm id}_{\rho'}\right).\label{iPR} \ee
Vaguely speaking, this condition just means that only the equivalence classes of the representations matter.

Next,  given any $(\kappa,\rho)$ on the left hand side of (\ref{P}), we can use the equivalence relation to reorient the faces $f$ containing $e$, such that their
orientations agree with that of $e$, and therefore an operator $P_e$ should be a map
\be P_e\ :\bigotimes_{f\supset e}\Hil_f \rightarrow  \bigotimes_{f\supset e}\Hil_f,\ee
and set
\be P_e\ =\ P_{R_e} \ee
with the unordered sequence  $R_e$ of the representations $\rho_f$ where $f$
ranges the set of faces containing $e$.

\subsubsection{A general solution for the conditions defining natural models}
It is not hard to see, that the  set of conditions defining  the class of the natural operator spin foam models has a general solution.

First, the assumed consistency with the face splitting equivalence of Sec \ref{subsubsectERFS} implies that
\be P_{R}\ =\ {\rm id} \ee
for every unordered sequence  $R$ given by the pair of elements $\rho$ and
$\rho^*$. Secondly, the consequence of the edge splitting equivalence of Sec
\ref{subsubsecERES}
is, that for every unordered sequence $R$ of irreducible representations, the operator $P_R$ (\ref{PR}) satisfies
\be P_RP_R\ =\ P_R. \ee
Hence, each operator $P_e$ is an orthogonal projection
onto a subspace
\be \Hil^{\rm s}_R\subset \Hil_R.\label{HsR}\ee
The subspaces $\Hil_R^{\rm s}$ are subject to the isomorphisms following
from (\ref{iPR}). They give rise to subspaces $\Hil_e^{\rm s}$ assigned to the internal edges $e$ of the 2-complexes
\subsection{Examples}
In the
following, we will show how different choices of the  operator labelling $P$,
defining different operator spin foam models,
reproduce different state-sum models. All the examples we discuss below, fall into the class of the natural operator spin foam models. Hence, by construction, each operator (\ref{Pe}) is a projection. The freedom consists
in fixing a subspace (\ref{HsR}),
\be \Hil_R^{\rm s}\ \subset\ \Hil_R\ =\ {\rm Inv}\bigotimes_{\rho\in R}\Hil_\rho\ee
for every unordered sequence $R$  of the equivalence classes of irreducible representations of $G$ (see the conditions (\ref{iPR})).

\subsubsection{Surjective $P$: BF theory}
 The easiest nontrivial choice is, of course, choosing (\ref{Pe}) $P_e$ to be the identity, for every edge $e$,
 \be P_e={\rm id} :\Hil_e\rightarrow\Hil_e,\ee
that is the fixed Hilbert subspace for each unordered sequence  $R$
of the irreducible representations is the full Hilbert space of invariants,
\be \Hil_R^{\rm s}\ =\ \Hil_R. \ee

Within this model, consider all the possible operator spin foams $(\kappa,\rho,P)$ defined on a fixed 2-complex $\kappa$ without boundary. Notice, that in the boundary free case,  the operator spin foam operator ${\cal Z}(\kappa,\rho,P)$  of
(\ref{Z}) is a $\mathbb{C}$-number.

For this choice of $P$, it is shown in \cite{OEPF} that, for any set of square-integrable functions
$$\{S_f:G\to \mathbb{C}\ :\ f\in\kappa^{(2)}\}$$
one has\footnote{Strictly speaking, the calculation in \cite{OEPF} is done on a two-complex consisting of the edges and faces of a hypercubic lattice. However, it is straightforward to generalise the calculation to arbitrary two-complexes.} that
\begin{eqnarray}\label{Gl:HendryksAndRobertsFormula}
\int_{G^E}\left(\prod_{e}d h_e\right)\prod_fS_f\left(g_f\right)\;=\;\sum_{\rho}\left(\prod_f\hat S_f(\rho_f)\right)\,{\cal Z}(\kappa, \rho, P)
\end{eqnarray}
\noindent where $e$ ranges  the set of edges $\kappa^{(1)}$, $E=|\kappa^{(1)}|$,
$f$ runs through the set of faces $\kappa^{(2)}$,
$$g_f:=\prod_{e\subset f}h_e$$
is the holonomy around a face $f$, and
$$\hat S_f(\rho)=\frac{1}{{\rm dim}\rho}\int_Gdg\,S_f(g)\chi_{\rho}(g)$$
is the Fourier coefficient of $S_f$.
In the formal limit of all $S_f$ approaching the delta function of $G$, one has $\hat S_f\equiv 1$, and the right hand side of (\ref{Gl:HendryksAndRobertsFormula}) approaches the BF-theory amplitude, e.g. the Ponzano-Regge amplitude \cite{PR} if $G=SU(2)$ and $\kappa$ is dual to a triangulation of a $3D$ manifold.

\subsubsection{Rank-one-$P_e$: The Barrett-Crane model}
The next model on the list of easy nontrivial examples, is the case when for every edge $e$ of each operator spin foam $(\kappa,\rho,P)$ of a model, the rank
of the projection operator $P_e$ is either $0$ or $1$. In fact, an example of a model of this type has been introduced by Barrett-Crane. In terms of our framework
it is a $G=Spin(4)\sim SU(2)\times SU(2)$ operator spin foam model. The representations associated to the faces of $(\kappa,\rho,P)$ are therefore $$\rho_f=(\rho_{j_f^+},\rho_{j_f^-}),$$
where $j_f^\pm$ are half-integers labelling
the $SU(2)$ representations, which -- in the picture of Euclidean $4D$ gravity -- constitute the self-dual and anti-self-dual part of the $Spin(4)$-connection. The projector $P_e$ assigned to each edge $e$ is zero,
\be P_e\ =\ 0,  \ee
unless every representation associated to a face $f$ hinging on the edge $e$ is \emph{balanced}, i.e. satisfies
$$j_f^+ = j_f^-\equiv j_f.$$
In the latter case, there is defined a unique element $\iota_{BC}\in \mathcal{H}_{e}$, called the "Barrett-Crane-intertwiner", and $P_e$ is set to be
\begin{equation}
P_e\;=\;\iota_{e{\rm BC}}\otimes\iota_{e{\rm BC}}^\dagger\,.
\end{equation}
In the balanced case (below ${\rm Inv}_{\rm SU(2)}\dots$ stands for the subspace of the SU(2) invariants; the subscript appears because we are dealing also with the Spin(4) group),
\be \Hil_e\ =\ {\rm Inv}_{\rm SU(2)}\left(\bigotimes_{f:e\subset f}\Hil_{j_f}\right)\,\otimes\,{\rm Inv}_{\rm SU(2)}\left(\bigotimes_{f:e\subset f}\Hil_{j_f}\right)\ee
where $\Hil_{j_f}$ is the carrier Hilbert space of the corresponding $SU(2)$ representation.
\noindent The Barrett-Crane intertwiner is the bilinear form
defined in the Hilbert space ${\rm Inv}_{\rm SU(2)}\left(\bigotimes_{f:e\subset f}\Hil_{j_f}\right)^*$ by the restriction of the canonical invariant bilinear form
defined in $\bigotimes_{f:e\subset f}\Hil_{j_f}^*$.

It can be constructed as follows: denote by $\epsilon_j\in\mathcal{H}_{j}\otimes\mathcal{H}_{j}$ the unique up to rescaling  SU(2) invariant. Furthermore, denote by
$$\pi:\bigotimes_{f:e\subset f}\Hil_f\ \to\ \mathcal{H}_e$$
the orthogonal projector. The Barrett-Crane intertwiner is then given by
\begin{eqnarray}
\iota_{BC}\;=\;c\;\pi(\bigotimes_{f:e\in f}\epsilon_{j_{f}})
\end{eqnarray}

\noindent where $c$ is a constant chosen such that $\iota_{BC}$ is normalised.

\subsubsection{Lessons from the previous two examples}
The previous two examples give us an interpretation of the natural operator spin foam models. Each natural $G$ operator spin foam model can be thought of as the $G$ BF
theory with constraints. Given an operator spin foam $(\kappa,\rho,P)$ of a given model,  elements of the Hilbert subspaces $\Hil^{\rm s}_e$ (ref{}) assigned to the edges are quantum solutions to the constraints. In the case of the Barrett-Crane model,
the constraint is intertwining the operators defined in
$\bigotimes_{f}\Hil_{j_f^+}$, and, respectively, in $\bigotimes_{f}\Hil_{j_f^-}$,
and the Barrett-Crane solution is the identity map, provided the representations are balanced.

\subsubsection{The natural operator spin foam model for the EPRL intertwiners}
The EPRL model \cite{EPRL} was developed to overcome some of the difficulties
one was encountering with the attempt to interpret the Barrett-Crane model as a
state-sum model for $4D$ Euclidean gravity. The fact that the operator labelling
 for the Barrett-Crane model assigns to the edges of the foams  (at most) rank
one operators lead to the argument that the theory does not capture enough
degrees of freedom (and in particular is not compatible with an LQG boundary
Hilbert space) \cite{graviton}.

In the EPRL model, again $G=SU(2)\times SU(2)$.\footnote{There is -- as well as for the Barrett-Crane model -- a Lorentzian version available \cite{BCLORENTZ, ASYMP1}, which uses different symmetry groups, but which are not discussed in this article.}
 Similarly, the projector $P_e$, for every edge $e$ of an operator spin foam $(\kappa,\rho,P)$, is defined by specifying its image, that is the corresponding subspace $\Hil^s_R$ of (\ref{HsR}). The EPRL model relies on the so-called "Barbero-Immirzi parameter" $\gamma$, which needs to be a real number $\gamma\neq 0,\pm 1$. The EPRL model subspace $\Hil^{\rm s}_e$
denoted here by $\Hil^{\rm s,EPRL}_e$ is nonempty only if, for every face, there is a half-integer $k_f$ such that
\be j_f^\pm=\frac{1}{2}|1\pm\gamma|k_f\label{kf}\ee
are also half-integers. The elements of this space ${\cal H}_e^{\rm s,EPRL}$ are called "EPRL intertwiners".
In \cite{EPRL} the EPRL map
\begin{eqnarray}
\iota^{\rm EPRL}_{ \gamma }\;:\;{\rm Inv}_{SU(2)}\left(\rho_{k_1}\otimes\ldots\otimes\rho_{k_n}\right)\;\longrightarrow\;
{\rm Inv}\left(\rho_{(j_1^+,j_1^-)}\otimes\ldots\otimes\rho_{(j_n^+,j_n^-)}\right)
\end{eqnarray}
is defined for any unordered sequences of admissible half integers
$$\tilde{R}\ =\ (k_1,...,k_n), \ \ \ R\ =\ ((j^-_1,j^+_1),...,(j^-_n,j^+_n))$$
\noindent which maps $SU(2)$-intertwiners $\eta$ to EPRL intertwiners $\iota^{\rm EPRL}_\gamma(\eta)$. The space ${\cal H}^{\rm s,EPRL}_R$ of the EPRL intertwiners
is therefore the image of the map $\iota_\gamma^{\rm EPRL}$, which can be shown to be one-to-one \cite{SFLQG}, but not an isometry, i.e. it does not preserve the Hilbert space inner product \cite{cEPRL}. Using this map, one maps a  (typically orthonormal) basis
$$\tilde{\cal B}\ \subset\ {\rm Inv}_{SU(2)}\left(\rho_{k_1}\otimes\ldots\otimes\rho_{k_n}\right)$$
into a basis
$${\cal B}^{\rm EPRL}\subset \Hil^{\rm s,EPRL}_R$$
(typically not orthonormal). In this way, for every edge $e$, the corresponding  subspace ${\cal H}_e^{\rm s,EPRL}\subset \Hil_e$ is equipped with a basis ${\cal B}^{\rm EPRL}_e\subset \Hil^{\rm s,EPRL}_e$, elements of which are $\iota^{\rm EPRL}_e(\eta_e)$, where $\eta_e$ ranges through a basis ${\tilde B}_e$ of the corresponding  space (via  (\ref{kf})) $\Hil_e^{\rm SU(2)}$ of the $SU(2)$ intertwiners.
We can expand the operator $P_e$ in the basis ${\cal B}^{\rm EPRL}_e$:
\be
 P_e=\sum_{\eta_e,\eta'_e} P^{\eta_{e}}_{\eta'_{e}} \iota^{\rm EPRL}_{ \gamma
}(\eta_{e}) \otimes \left(\iota^{{\rm EPRL}}_{\gamma}(\eta'_{e})\right)^\dagger,
\ee
where the coefficients $P^{\eta_{e}}_{\eta'_{e}}$ are defined by
the Hilbert product $\left(\cdot|\cdot\right)_e$ in $\Hil_e$, namely
\be \sum_{\eta'_e} P^{\eta_{e}}_{\eta'_{e}} \left(\iota^{\rm EPRL}_{ \gamma
}(\eta'_e)|\iota^{\rm EPRL}_{ \gamma
}(\eta''_e)\right)_e\ =\ \delta^{\eta_{e}}_{\eta''_e}\,. \ \ee
\begin{figure}[hbt!]
  \centering
\includegraphics[width=0.4\textwidth]{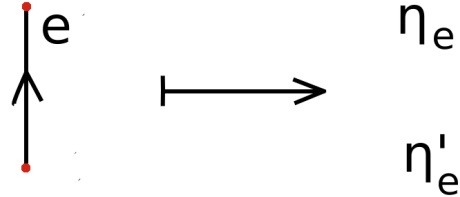}
\caption{Two $SU(2)$ intertwiners $\eta_e,\eta'_e$ are assigned to the end and,
respectively, the beginning point of each edge $e$}
\label{colour_EPRL}
\end{figure}
As a result, given an operator spin foam $(\kappa,\rho,P)$,  instead of assigning an operator $P_e$ to each edge $e$, one considers a set of assignments $\eta$ of two
$SU(2)$ intertwiners $\eta_e,\eta'_e\in \Hil^{\rm SU(2)}_e$,  to the end and,
respectively, the beginning point of each edge $e$ (FIG. \ref{colour_EPRL}). 
Following the derivation of the amplitude
form of the partition function done
in \cite{cEPRL} we obtain for the case of a oriented 2-complex with boundary:
\be
\begin{split}
{\cal Z}(\kappa,\rho,P)\ = \sum_{\eta}\prod_e P^{\eta_e}_{\eta'_e}&\prod_{f}(2j^+_f+1)(2j^-_f+1)\prod_{v}A_v\\
&\prod_{\tilde{l}} \frac{1}{\sqrt{(2j^+_{f_{\tilde{l}}}+1)
(2j^-_{f_{\tilde{l}}}+1)}}
\bigotimes_{{\tilde e}}\iota^{\rm EPRL}_{ \gamma
}(\eta_{e})   \otimes \bigotimes_{{\tilde e}'}
\left(\iota^{{\rm EPRL}}_{\gamma}(\eta'_{e'})\right)^\dagger
\end{split}
\label{statesum'}
\ee
where
$f$ runs through the set of faces, $v$ ranges the set of the internal vertices, $l$ ranges the set of the boundary edges (links) and $f_{l}$ is the (unique) face containing ${l}$, and ${\tilde e}$/${\tilde e}'$ ranges the set of edges which intersect $\partial \kappa$ at the end/beginning point, and  $A_v$ is the vertex amplitude (\ref{Av}).

Note that the $P^{\eta_e}_{\eta'_e}$ matrix is not appearing in the original definition of the EPRL state sum in \cite{EPRL}. It has to be included if $P_e$ is supposed to be an orthogonal projection, since the EPRL map $\iota_\gamma^{EPRL}$ is not an isometry. The $P^{\eta_e \eta_{(w,e)}}$ can be interpreted as measure factor appearing when summing over intertwiners.  If the $P^{\eta_e}_{\eta'_e}$  factors are not included in the partition function, then the EPRL-intertwiners are summed over with a different measure, and lead to $P_e$ not being an orthogonal projection -- in particular, the operator ${\cal Z}(\kappa,\rho,P)$ is no longer invariant under trivially subdividing an edge.  

\section{Summary} The operator spin foams we have introduced are linear
combinations
of the usual spin foams, therefore they should be robust in any spin foam
context. The first three ``moves'' defining the equivalence relation we
have constructed: reorientation of faces, edges, and splitting a face are
consequence of analogous moves and equivalence of the spin networks. The
equivalence upon splitting an edge
and the suitable relation between the operators is a choice natural for
the consistency between combining the operator spin foams and combining
the corresponding operators. Also the contraction  as well as the
operator spin foam  {\it operator} are naturally defined operations, that
exist in dependently on our believes and can be used as tools of any spin foam theory. The family of natural spin foam models we derived from assumed symmetry 
took appearance of constrained BF spin foam models. Each of the is defined by
the restriction of a proper spin foam model to a subspace in the space of
intertwiners. Since gravity is often viewed in that way, one of the
natural $Spin(4)$ operator spin foam model characterised by suitable
subspace of solutions to the simplicity constraints could be the proper
quantum gravity model. The most important example is given by the EPRL
subspace of the $Spin(4)$ intertwiners. In that case, the corresponding
natural operator spin foam model coincides with the proposal of
\cite{cEPRL}, whereas it is different than the EPRL proposal \cite{EPRL}.
That difference was already emphasised in \cite{cEPRL}. The new conclusion
coming from the current work is the set of rules governing operator spin foams that
is satisfied in one case and is not satisfied by the other one. 
If experiment shows that nature favours the less natural model, we should still
understand better its operator structure.   
\

\noindent{\bf Acknowledgments} 
We have benefited a lot from stimulating discussions with Carlo Rovelli, Abhay
Ashtekar, Jonathan Engle, Claudio Perini and Helena Perini.  
WK,MK,JL were partially supported by: (i) the grants N N202 104838, N N202
287538 and 182/N-QGG/2008/0 (PMN) of Polish Ministerstwo Nauki i Szkolnictwa Wyzszego; (ii)
the grant Mistrz of the Fundation for the Polish Science. JL also acknowledges the US National Science Foundation (NSF) grant PHY-0456913.
All the authors thank the ESF  network ’Quantum Geometry and Quantum Gravity’ for short visit grants to collaborate.

\end{document}